Network Characteristics Control Social Dilemmas in a Public Good Game:

A Mechanical Framework

Authors: Chulwook Park[1]

[1] International Institute for Applied Systems Analysis (IIASA), A-2361 Laxenburg, Austria.

Correspondence should be addressed to Chulwook Park; parkc@iiasa.ac.at

## Abstract

**Purpose:** We propose a model to present a possible mechanism for obtaining sizeable behavioral structures by simulating an agent based on the evolutionary public good game with available social learning. **Methods:** The model considered a population with a fixed number of players. For each round, the chosen players may contribute part of their value to a common pool. Then each player may imitate the strategy of another player, based on relative payoffs (whoever has the lower payoff adopts the strategy of the other player) and change strategy using different exploratory variables. **Results:** Relative payoffs are subject to incentive, including participation costs, but are also subject to mutation, the rate of which is sensitized by network characteristics (social ties). **Values:** The process covered by this study is of interest and is relevant across a broad range of disciplines that use game theory, including the framework of cultural evolutionary dynamics.

# Background

Many computational simulations in game theory have shown that a substantial fraction of players are willing to invest costs (i.e., contribute to a joint effort) to increase their fitness (Rockenbach, & Milinsk, 2006). By and large, the rate of incentive (reward) looms credibly enough to increase the average level of pro-social contributions. Indeed, the system is itself a public good, and the players are often seen as altruistic because others benefit from their costly efforts (Fowler, 2005).

Conversely, there are always players who take advantage of other's incentive toward the public good called free-riders (or defectors). Free-riding should spread among voluntary players and ultimately cause the cascade of the system. A solution to protect this system against defection is to refrain the free-riders (Boyd, & Richerson, 1992) because their strategy is taken up other players in turn, leading to infinite regress. Moreover, if everyone contributes to the public good, risk will propagate without need of an incentive. Their number of free-rider may grow through simple cultural evolution, ultimately allowing the defector to invade with impunity. We show how a simple step-by-step mechanism can overcome this objection, using one of the most common properties, called social learning (imitation, exploration) and social ties in a random network.

## Model

**Games necessary (payoff calculation):** The term game here describes a set of interactions among a set of players. The players act according to their behavioral phenotypes, called strategies. The moves of the game are decisions stemming from interactions between two (or more) co-players with different views of strategy, which translate into payoffs.

$$\begin{pmatrix} \alpha & \beta \\ \gamma & \delta \end{pmatrix}$$

Most conceptually straightforward games offer only two strategies for each player (up or down and left or right) and four outcomes ($\alpha, \beta, \gamma, \delta$), and they generally involve two players. However, a major challenge, even in the simple game, is to be determining how to rank payoff values. This model was able to achieve this by simply defining the dynamics of two strategies with four outcomes. Suppose that players must choose between $n$ options or strategies. In a simple case, player one and player two simultaneously reveal a penny. If both pennies show heads or both show tails, player two must pay player one \$1. On the other hand, if one penny shows heads and the other shows tails, player one must pay player two \$1. The game, then, can be described as follows;

$$Player\ 2$$

|  |  | $H$ | $T$ |
|---|---|---|---|
| $Player\ 1$ | $H$ | $(-1,1)$ | $(1,-1)$ |
| | $T$ | $(1,-1)$ | $(-1,1)$ |

The matrix describes, in the $i$th row and $j$th column, the pair $(a_{ij}, b_{ij})$ of payoff values. It shows that if the outcome is (-1, 1), player one, who here chooses the row of the payoff matrix, would have done better to choose the other row. On the other hand, if the outcome had been (1, -1), it is player two, the column player, who would have done better to switch. Players one and two have diametrically opposed interests here.

This is a very common strategic situation in games. During a game of soccer, in a penalty kick, if the striker and keeper are mismatched, then the striker is happy, but if they are matched, the keeper is happy. This logic applies to many common preference situations, and an efficient way to represent these dynamics is to let a player decide to implement a strategy with a given probability.

$$x = (x_1, x_2, \ldots, x_n), \qquad x_1 \geq 0\ and\ x_1 + \cdots + x_n = 1$$

Where the probability distribution presents two or more pure strategies, among which players choose randomly, denoted as the set of all probability distributions of such mixed strategies, in this way;

$$p_r(x_1) + p_r(x_2) + \cdots + p_r(x_n) = 1 = 100\%$$

$$p_r(x_i) \in [0,1]$$

For all events, the player expects to be happy half of the time and unhappy half of the time.

*Player* 2

|            |        | $H(.5)$   | $T(.5)$   |
|------------|--------|-----------|-----------|
| *Player* 1 | $H(.5)$ | $(-1,1)$ | $(1,-1)$  |
|            | $T(.5)$ | $(1,-1)$ | $(-1,1)$  |

It does not matter what the players do because they cannot change the outcome, so they are just as happy to flip a coin (in the coin game) or choose one direction of two (in soccer). The only question is whether one is able to anticipate the decision of the other player. If striker knows that keeper is playing heads (one direction of two), the striker will avoid heads and play tails (the other direction of the two). However, we cannot usually guess the coin flipping if we change the the payoff to provide a different outcomes. As a matter of facts, if the game expands to the following arbitrarily mixed payoff conditions $\{A > B > C\}$,

*Player* 2

|            |        | *Left*   | *Right*  |
|------------|--------|----------|----------|
| *Player* 1 | *Up*   | $(B,A)$  | $(C,C)$  |
|            | *Down* | $(C,C)$  | $(A,B)$  |

A player's expected utility is as follows:

$$\sigma u = \frac{b-c}{a+b-2c}, \qquad a+b-2c > 0$$

$$\sigma u = \frac{b-c}{a+b-2c}, \qquad b-c > 0, \qquad a+b-2c > 0$$

$$\sigma u = \frac{b-c}{a+b-2c}, \qquad a+b-2c > b-c, \qquad a > c$$

Even in the slightly different payoff related to arbitrarily given mixed-conditions $\{A > B > C\}$, the payoff will always be positive for [0,1] in the above conditions.

Now, with this dynamic, if we determine the probability that each outcome occurs for some percentage of instances, the payoff of players one and two appear below:

$$Left(\tfrac{2}{3}) \qquad Right(\tfrac{1}{3})$$

$$Up(\tfrac{1}{3}) \quad (B=1, A=2) \quad (C=0, C=0)$$

$$Down(\tfrac{2}{3}) \quad (C=0, C=0) \quad (A=2, B=1)$$

The probability of each outcome must be multiplied by a particular player's payoff.

$$
\begin{array}{c}
\quad\quad 2/3 \quad\quad\quad\quad 1/3 \\
\begin{array}{cc}
1/3 & \begin{pmatrix} B=1, A=2 \\ \dfrac{1}{3}*\dfrac{2}{3}=\dfrac{2}{9} \end{pmatrix} \begin{pmatrix} C=0, C=0 \\ \dfrac{1}{3}*\dfrac{1}{3}=\dfrac{1}{9} \end{pmatrix} \\
2/3 & \begin{pmatrix} C=0, C=0 \\ \dfrac{2}{3}*\dfrac{2}{3}=\dfrac{4}{9} \end{pmatrix} \begin{pmatrix} A=2, B=1 \\ \dfrac{2}{3}*\dfrac{1}{3}=\dfrac{2}{9} \end{pmatrix}
\end{array}
\end{array}
$$

Then, all those numbers are summed together, giving player one's payoff as follows:

$$\frac{2}{3}=\frac{6}{9}=\frac{2}{9}+0+0+\frac{4}{9}=1\left(\frac{2}{9}\right)+0\left(\frac{1}{9}\right)+0\left(\frac{4}{9}\right)+2\left(\frac{2}{9}\right)$$

Player one's earned mixed-equilibrium is {2/3}. Then, player two's payoff is as follows:

$$\frac{2}{3}=\frac{6}{9}=\frac{4}{9}+0+0+\frac{2}{9}=2\left(\frac{2}{9}\right)+0\left(\frac{1}{9}\right)+0\left(\frac{4}{9}\right)+1\left(\frac{2}{9}\right)$$

This means that player two's earned mixed-equilibrium is also {2/3}. Checking for the underlying assumption of this game's probability distribution $p_r(x_i) \in [0,1]$, it is clear that all of these payoffs will always be positive for [0,1] because they all must added in the following way:

$$\frac{2}{9}+\frac{1}{9}+\frac{4}{9}+\frac{2}{9}=1$$

These also satisfy the rules of probability distributions:

$$p_r(x_1)+p_r(x_2)+\cdots+p_r(x_n)=1=100\%$$

This relative welfare distribution derived directly from two by two games will allow us to set a fundamental assumption for evolutionary dynamics.

**Replicator dynamics:** Now, let the model consider a population of players, each with a given strategy. From time to time, the players meet randomly and play the game according to a plan. If we suppose that each player is rational, individuals consider several types of different payoffs for each:

$$x_i = \Pr(i)\pi(i)$$

Where each palyer has the payoff $\{\pi(i)\}$, which shows how well that type $(i)$ is doing, and each type has a proportion $\{\Pr(i)\}$. Then, they choose a certain strategy they consider to be the best outcome from the entire population.

$$\dot{x}_i = \frac{P_r(i)\pi(i)}{\sum_{j=1}^{N} P_r(j)\pi(j)}$$

Where $P_r(i)$ is the proportion of the different types, and $\pi(j)$ is the payoff for all of the types. Here, we consider a population of types, where those populations are succeeding at various levels. Some are doing well, and some are not. The dynamics of the model suppose a series of change in distribution across types, such that there are set of types $\{1,2, \ldots, N\}$, a payoff for each type $(\pi_i)$, with a proportion of each as well $(Pr_i)$. The individual player's a strategy $(\dot{x}_i)$ in each round is given a probability that is the ratio of this weighting to the all possible strategies, where, $\dot{x}_i$ is the probability that an individual play will take a strategy times payoff $[P_r(i)\pi(i)]$, divided the sum of the weightings of all the different strategies $[\sum_{j=1}^{N} P_r(j)\pi(j)]$.

Thus, the probability that the individual player will act in a certain way in the next round is just that action's relative weighting. Specifically, let's propose that there are different probabilities $P_r(i)$ of using different strategies $(x, y, and\ z)$: that is, strategy $x$ has a probability of 40%, strategy $y$ also has a probability of 40%, and strategy $z$ has a 20% probability. That might lead us to guess the strategy $x$ and $y$ are better than $z$. However, we could also look at the payoffs $\pi(i)$ of the different strategies. For instance, it might be that using strategy $x$, we can obtain a payoff 5, using strategy $y$ there is a payoff 4, and using strategy $z$, players take payoff 6. This prompts us to consider what strategy we should use, and the answer depends on both the payoff and the probability.

In this dynamic, the model presented here proposes a description of how individuals might choose what to do or which strategies are best. Because after certain move, some will appear to be doing better than others, the ones doing worst are likely to copy the ones doing better. Based on the cultural evolutionary assumptions for public goods game, we specified the respective frequencies of actions of cooperators ($P_c$), defectors ($P_d$), and loners ($P_l$). The experimenter assigns a value to each players; then, the players may contribute part (or all) of their value to the game (a common pool).

| | | |
|---|---|---|
| Cooperators, ready to join to the group and to contirbute to its effort | $P_c$ | n.a. |
| Defectors, who join but do not contribute | $P_d$ | n.a. |
| Loner, unwilling to join the public goods game | $P_l$ | n.a. |

**Table 1.** Strategies of potential players.

In each round, a sample ($S$) of individuals are chosen randomly from the entire population $N$. $S$ ($0 \leq S \leq N$) out of the individuals participate in the game, paying a cost ($g$). Each round requires at least two (cooperator and defector) participants, and others must be nonparticipants. The cooperators contribute a fixed amount of value $c > 0$ and share the outcomes multiplied by the interest rate $r$ ($1 > r > N$) equally among all others $S - 1$ participants, defectors are in the round but do not contribute values. During the round, the payoffs for the strategies $P_c$, $P_d$, and $P_l$ (denoted by $x, y,$ and $z$) are then determined, with a participation cost $c$ and an interest rate $rc$, based on the relative frequencies of the strategies ($n_c/N$).

$$P_c = -c + rc\frac{n_c}{N}, \qquad P_d = rc\frac{n_c}{N}, \qquad 1 < r < N$$

Where $P_c$ is the payoff of the cooperators, $P_d$ is the payoff of the defectors, $rc$ is the interest rate ($r$) multiplied by fixed contribution cost ($c$) for the common good. For the expected payoff values of the cooperators ($P_C$) and defectors ($P_D$), a defector in a group with $S - 1$ coplayers ($S = 2, \dots, n$) obtains from the common good a payoff of $rcx/(1 - z)$ on average because the nonparticipants ($z$) have a payoff of 0 (Sasaki et al., 2012).

$$P_D = \left(rc\frac{x}{1 - z} - g\right)(1 - z^{n-1})$$

Where, $z^{n-1}$ is the probability of finding no coplayer. In the abstract, we can write this as the derivative of $z$ with respect to $n$ for any power $n$, $nz^{(n-1)}$. This is known as the power rule and is symbolically represented as the exponent (i.e., $z^2 = 2z^1$, $z^3 = 3z^2$, ..., $z^n = nz^{n-1}$). This is what it means for the derivative of $z^n$ to be $nz^{n-1}$ and for the population thus to be reduced to loners (nonparticipation). In addition, cooperators contribute effort $c$ with a probability $1 - z^{n-1}$. Hence,

$$P_D - P_C = c(1 - z^{n-1})$$

The average payoff ($\bar{P}$) in the population is then given in this way:

$$\bar{P} = (1 - z^{n-1})[(r - 1)cx - (1 - z)g]$$

The replicator equation gives the evolution of the three strategies in the population:

$$\dot{x} = x(P_x - \bar{P}), \qquad \dot{y} = y(P_y - \bar{P}), \qquad \dot{z} = z(P_z - \bar{P})$$

The frequencies $x_i$ of the strategies $i$ can simply be represented by

$$\dot{x}_i = x_i(P_i - \bar{P})$$

Where $x_i$ denotes the frequency of strategy $i$, $P_i$ is the payoff of strategy $i$, and $\bar{P}$ represents the average payoff in the population. Accordingly, a strategy will spread or dwindle depending on whether it does better or worse than average. This equation holds that populations can evolve, in the sense that the frequencies $x_i$ of strategies change with time. Thus, we let the state $x(t)$ depend on time and denote this by $\dot{x}_i(t)$, where $\dot{x}_i$ changes $dx_i/dt$.

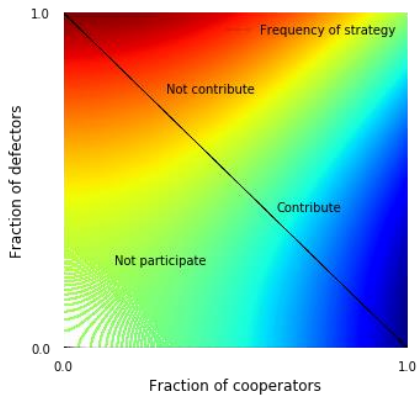

**Figure 1.** Schematic illustration of the model (PGG) with replicator dynamics. Colars indicate prototype as their proportion of the implementation by cooperators (blue), defectors (red), and loners (yellow to green). The dotted line denotes their relative frequencies, where defectors dominate cooperators, loners dominate defectors, and cooperators dominate loners.

This model is particularly interested in the growth rates of the relative frequencies of the strategies. In other words, the state of the population evolves according to the replicator equation, where the growth rate ($\dot{x}_i / x_i$) of a strategy's frequency corresponds to the difference between its payoff ($P_i$) and the average payoff ($\bar{P}$) in the population.

**Imitation dynamics with updating algorithm:** In the cultural evolutionary context we are considering here, strategies are unlikely to be inherited, but they can be transmitted through social learning (Avital, & Jablonka, 1994). If we assume that individuals imitate each other, replicator dynamics will be yielded again. To be more specific, from time to time, a randomly chosen individual from the population imitates a given model with a certain likelihood. Thus, the probability that an individual switchs from one strategy to the other is given as

$$p = x_i \sum_j \left( P_{ij} - P_{ji} \right) x_j$$

This is simply the replicator equation, but it describes that a player ($P_i$) making a comparison with another ($P_j$) player will adopt the other's strategy only if it promises a higher payoff. This switch is more likely if the difference in payoff is a function of the frequencies of all strategies, based on pairwise interactions (Traulsen, & Hauert, 2009): The focal individual compares its payoff ($P_i = \pi_f$) with the payoff of the role individual ($P_j = \pi_r$), and then the focal individual chooses to imitate (or not) the role individual given the following:

$$p = \left[ 1 + e^{-\beta(\pi_r - \pi_f)} \right]^{-1}, \qquad \beta = selection\ intensity$$

This mechanism holds the factorial for payoff, that is, how many combinations of $r$ objects can we take from $n$ objects:

$$n! = \prod_{k=1}^{n} k, \ \rightarrow \ nCr = n!/r!\,(n-r)!$$

with the Gillespie algorithm (stochastic dynamic) for updating the system ($a_{r-1} / a_{tot} < z_1 < a_r / a_{tot}$).

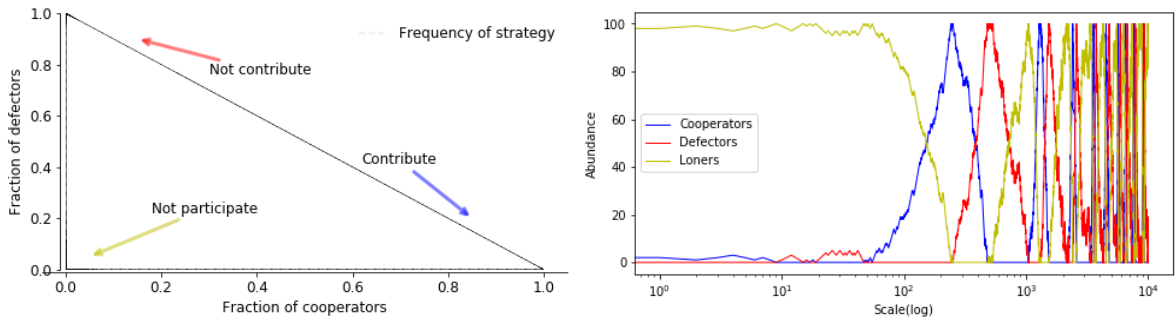

**Figure 2.1.** Simulation 1 (prototype) with the replicator dynamics: The plot on the left-hand side shows a well-mixed population with a finite number of strategies [cooperators (blue arrow), defectors (red arrow), and loners (yellow arrow)] with incentive. The parameters are as follows: N = 5, c = 1, u = 1e-10, g = 0.5, and interest rate r = 3. The plot on the right-hand side shows the categorized oscillation of the strategies (cooperators = blue, defectors = red, and loners = yellow).

| Cooperator | | C | n.a. |
|---|---|---|---|
| Defector | | D | n.a. |
| Loner | | L | n.a. |

**Table 2.1.** Strategies of potential players (C: cooperation; D: defection; L: no participation).

| Parameters | Number of individuals | M | 100 |
|---|---|---|---|
| | Number of samples | N | 5 |
| | Rounds per generation | tt | 1 |
| | Number of generations | t | 10,000 |
| | Participation cost | g | 0.5 |
| | Investment of participation | c | 1 |
| | Participation benefit | r | 3 |
| | Mutation rate | u | 1e-10 |

**Table 2.2.** Model variables, parameters, and default of the parameter values.

| Imitation dynamics | Selection intensity | s | 0 ~ 1 |
|---|---|---|---|
| | Imitation probability | pr | 0 ~ 1 |

**Table 2.3.** Imitation and exploration of the parameter values.

The above procedures assume a well-mixed population with a finite number of strategies that are proportional to its relative abundances given that the fitness values are frequency-dependent, coexisting in steady or fluctuating frequencies the evolutionary games (Figure 2.1). The mechanism is a combination of the rational process and the copying process, or in order words, an individual rationally chooses from a nearby individual because it seems that it would affect successful outcomes.

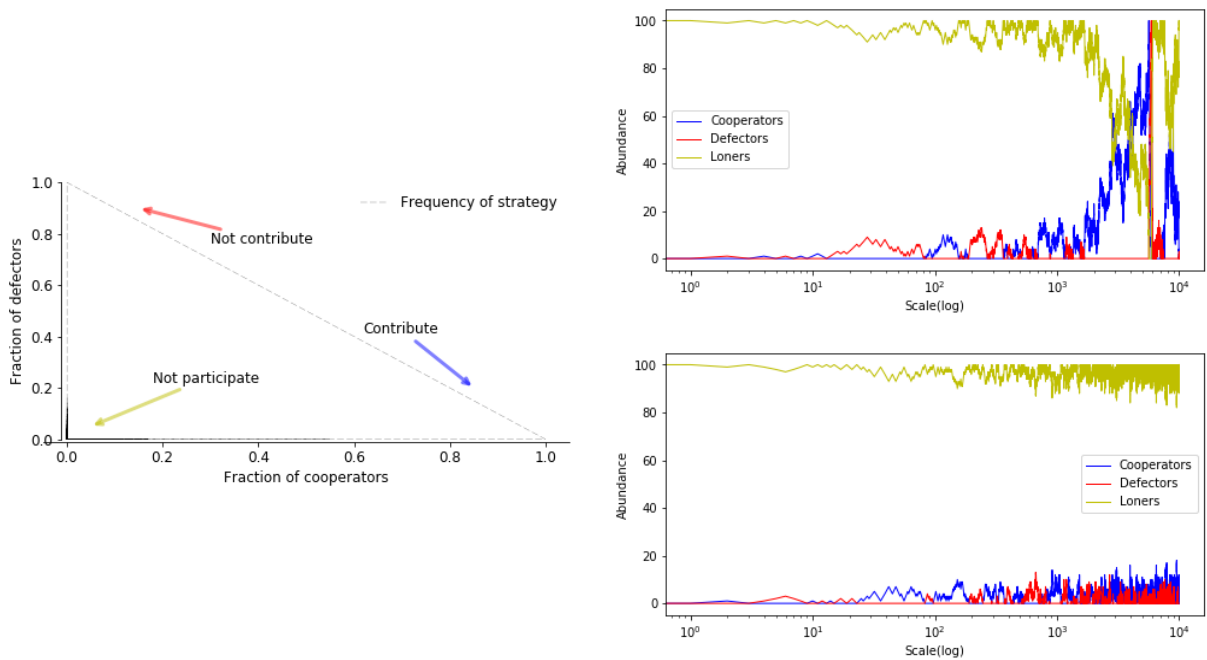

**Figure 2.2.** Replicator dynamics with parameters (participation cost). Plots of the left, and right upper are with participation cost (g): Parameters: n = 5, r = 3, c = 1, u = 1e-10, participation cost g = 2. The right bottom plot shows g = 3. Both plots on the right-hand side show the categorized oscillation of the strategies (cooperators = blue, defectors = red, and loners = yellow).

The simulation in Figure 2.2 shows that he system settles into different effects on the intermediate interest rate, with participation costs. As the interest rate increases $(r)$, it prompts the population to undergo stable oscillations relative to a global attractor, where the players participate by contributing to the public good. However, if that contribution is too expensive, that is, if the participation cost is $g \geq (r-1)c + l$ for rewarding, or $g \geq (r-1)c$ for punishing, the players will not opt to participate (Figure 2.3). In this scenario, nonparticipation becomes the global attractor (the right bottom plot on the Figure 2.2).

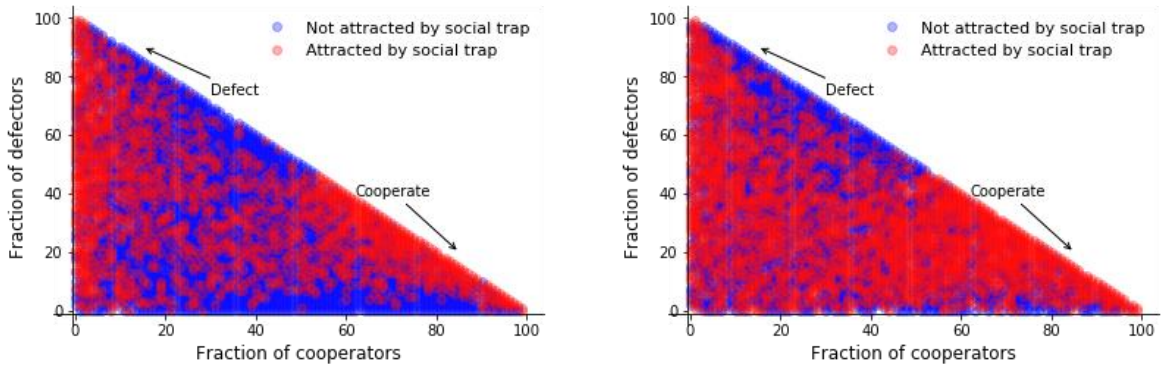

**Figure 2.3.** Cultural evolutionary dynamics with incentive. Parameters: N=5, c=1, u=1e-10, g=0.5, interest rate r = 3 (plot on the left), and r = 1.8. (plot on the right), with individuals attracted individuals by the social trap. Both plots show the categorized strategies (cooperators = blue, defectors = red).

**Replicator-mutator dynamics:** Not all learning is learning from others. We can also learn from our own experience. The dynamics of the replicator equation describes selection only, not drift or mutation. An intelligent player might adopt a strategy, even if no one else in the population is using it, if the strategy offers the highest payoff. Dynamics can also be modified with the addition of a small, steady rate of miscopying for any small linear contribution that exceeds the role of dynamics. As a result, the stability of the system changes, making the system structurally unstable. This can be interpreted as the exploration rate, and it corresponds to the mutation term in genetics (Sigmund, De Silva, Traulsen, & Hauert, 2010). Thus, by adding a mutation rate with a frequency-dependent selection, we expect that the impact of mutations can show a more general approach to evolutionary games, without explicit modelling of their origin (Nowak & Sigmund, 2004).

$$\frac{dx_i}{dt} = x_i(P_i - \bar{P}) + \mu(1 - x_i) - 2\mu x_i$$

In the set of model, both of those dynamics are in play, where individuals are both copying more prominent strategies and copying strategies that are doing better than the others. The fate of an additional strategy can be examined by considering the replicator dynamics in the augmented space (mutation) and computing the growth rate of the fitness that such types get in the case of evolution (Figure 3). The mechanism holds in the ordinary differential equation, namely, a differential equation containing one or more function *s* of an independent variable and its derivatives: $\frac{dy}{dx}, \frac{d^2y}{dx^2}, \ldots, \frac{d^ny}{dx^n}, \; x = independt \; variable, y = independt \; variable, and \; y = f(x) is \; an \; unknown \; function \; of \; x$] for updating the system .

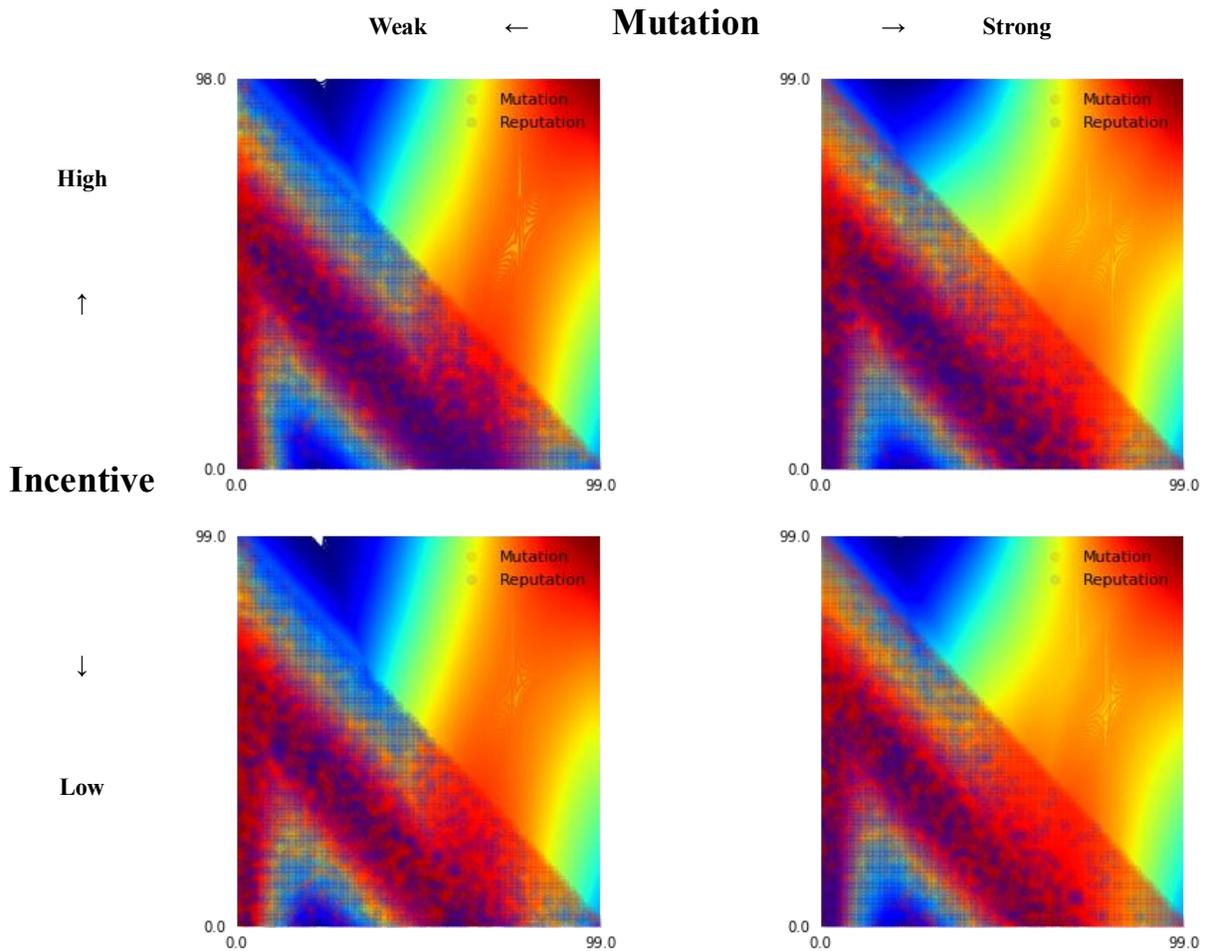

**Figure 3.** Cultural evolutionary dynamics with replicator-mutator dynamics. The parameters of the upper plots are n=5, c=1, g=0.5, r=3, u=1e-10 (left), and u=1e-1 (right). The bottom plots have n=5, c=1, g=0.5, r=1.8, and u=1e-10 (left), u=1e-1 (right). Colars indicate prototype as their proportion of the implementation by cooperators (blue), defectors (red), and loners (yellow to green).

| Cooperator | C | n.a. |
|---|---|---|
| Defector | D | n.a. |
| Loner | L | n.a. |

**Table 3.1.** Strategies of potential players (C: cooperation; D: defection; L: no participation).

| Parameters | Number of individuals | M | 100 |
|---|---|---|---|
| | Number of samples | N | 5 |
| | Rounds per generation | tt | 1 |
| | Number of generations | t | 10,000 |
| | Participation cost | g | 0.5 |
| | Investment of participation | c | 1 |
| | Participation benefit | r | 3 |
| | Mutation rate | u | 1e-10 |

**Table 3.2.** Model variables, parameters, and default of the parameter values.

| Imitation dynamics | Selection intensity | s | 0 ~ 1 |
|---|---|---|---|
| | Imitation probability | pr | 0 ~ 1 |
| | | | |
| Exploration dynamics | Exploration probability | pe | 0 ~ 1 |
| | Mean for normally increment | mu | 0 ~ 1 |
| | Standard deviation for normally increment | sigma | 0 ~ 1 |

**Table 3.3.** Imitation and exploration of the parameter values.

Figure 3 indicates that mutation has a significant effect on the transition of strategies. The system settles into the different effects of the intermediate mutation rate. As it decreases, the red individuals appear at lower rates (plot of the left side). This prompts players to participate by contributing to the public goods. If, on the other hand, as long as the mutation rate is high enough, nonparticipation becomes a global attractor; selfish players continually defect by refraining from contributing (right-hand plot).

**Replicator-mutator including network dynamics:** Current models currently proposed cannot explain cooperation in communists with different average numbers of social ties. To impose the social ties as a parameter, the primary feature of a random graph (Erdos & Rényi, 1959) was used for network characteristics, as follows;

First, individuals in the model considered as vertices (fundamental element drawn as a node), and sets of two elements are drawn as a line connecting two vertices (where lines are called edges) (Figure 4.1's left-hand side). Nodes are graph elements that store data, and edges are the connections between them, but the edges can store data as well. The edges between the nodes can describe any connection between individuals (called adjacency). The nodes can contain any amount of data with the information we decide to store in this application, and the edges include data on the strength of connections.

Networks have additional properties, in that edges can have a direction, meaning the relationship between two nodes only applies only in one direction, not the other. A directed network is the term for a network where shows a direction. In the present model, however, we used an undirected network, featuring edges with no sense of direction because with a network of individuals and edges that indicate two individuals who have met, directed edges might be unnecessary. Another essential property of this structure is connectivity. A disconnected network has some vertex (nodes) that cannot be reached the other vertices (Figure 4.1's right-hand side).

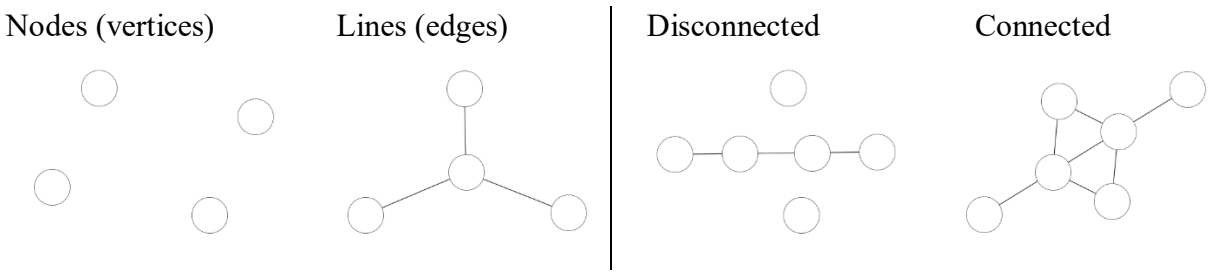

**Figure 4.1**: Schematic representation of the nodes and lines (left-had side), and connectivity (right-hand side).

A disconnected network might feature one vertex that is off to the side has no edges. It could also have two so-called connected components, which from a connected network on their own but have no connection between them. Thus, a connected network has no disconnected vertices. This could be a criterion for describing a network as a whole, called connectivity. The fulfillment of the criterion would depend on the information contained in the graph, usually controlled by (*n, p*).

An object-oriented language was used to allow the creation of vertex and edge objects and give each of them a property. A vertex could by identified by the list of edges that it is connected to, and the reverse would be true for edges. However, operations involving networks might be inconvenient if we must search through vertex and edge objects. Thus, we represent connections in networks that simply used a list of edges (Figure 4.2's left-hand side).

The edges themselves are each represented with an identifier of two elements (1,2). Those elements are usually numbers corresponding to the ID numbers of vertices. Thus, in the end, this list simply shows two nodes with an edge between them, and an edge list is a list that encompasses all smaller lists. Because the edge list contains a list of other lists, it is sometimes called a two-dimensional list. We represent the edge list in a network as an adjacency list. Our vertices normally exhibit ID number that corresponds to the index in an array (Figure 4.2's right-hand side).

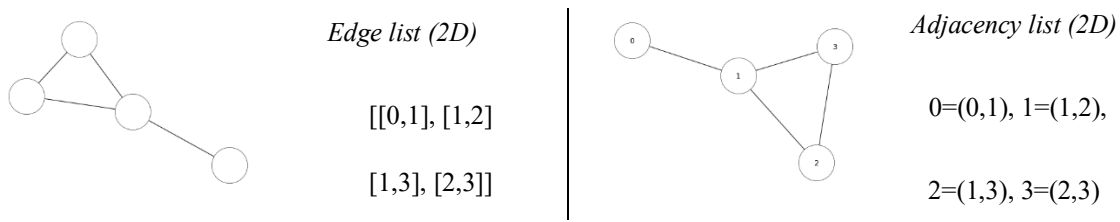

**Figure 4.2**: Schematic representation of the edge list (left-hand side) and the adjacency list (right-hand side).

In this array, each space is used to store a list of nodes, such that the node with a given ID is adjacent to the index with the same number. For instance, an opening at index 0 represents a vertex with an ID of 0. That vertex shares an edge with one node, so that reference to that node is stored in the first spot in the array. Thus, because the list contains other lists, the adjacency list is two dimensional allowing us to use an adjacency matrix that is essentially a two-dimensional array, but the lists within it are all the same length.

**Application (social ties)**: Due to the collection of nodes influenced by connection probabilities corresponding to the adjacency list, the distribution of the connection in the network was used for the social ties (probability of degree) as below:

$$\sum_{v \in V(G)} \deg(v) = \deg(v_1) + \cdots + \deg(v_n) = 2|E(G)|$$

In this context, one might wonder whether some groups of individuals interact with each other more and more often than with others and under which conditions social beings are willing to be cooperative. Moreover, they must be able to adjust their own changes to thrive. To understand a cooperative network of interaction, both the evolution of the network and the evolution of strategies within it should be considered simultaneously.

$$\frac{dx_i}{dt} = x_i(P_i - \bar{P}) + [\mu(1 - x_i) - 2\mu x_i] * \left[ \frac{\dfrac{2 - t}{N}}{\dfrac{N(N-1)}{2}} \right]$$

Where $t$ represent the social ties $[t = 2|E(G)|]$ between individuals, and $N$ is the nodes (Santos, Pacheco, & Lenaerts, 2006) of the sample of the population. $(2 - t)/N$ denotes the actual connection in the network $(AC)$, $N(N-1)/2$ denotes the potential connection in the network $(PC)$. A potential connection $(PC = N(N-1)/2)$ is a connection that could potentially exist between two individuals, regardless of whether or not it actually does. One individual could know another individual, and this object could connect to that one.

Whether the connection is actually there is irrelevant when we are talking about a potential connection. By contrast, an actual connection $(AC = (2 * t)/N)$ is one that actually exists ($t$ =social ties), where this individual does know that one, and this obejct is connected to that one. In relation to these small linear contributions and their dynamics, structural instability can be interpreted as the characteristics of the network, influenced by the exploration rate, which corresponds to the idea of mutation in genetics.

Grouping the network characteristics that incorporate the decisions of individuals through establishing new links or giving up existing links (Traulsen, Hauert, De Silva, Nowak, & Sigmund, 2009), we propose a version of evolutionary game theory and study the dynamical coevolution of individual strategies and network structure. In this model, the dynamics operate such that the population moves over time as a function of payoffs, and proportions based on the replicator-mutator dynamic multiplied its network density (Figure 5).

| | | |
|---|---|---|
| Cooperator | C | n.a. |
| Defector | D | n.a. |
| Loner | L | n.a. |

**Table 4.1.** Strategies of potential players (C: cooperation; D: defection; L: no participation).

| Parameters | | | |
|---|---|---|---|
| | Number of individuals | M | 100 |
| | Number of samples | N | 5 |
| | Rounds per generation | tt | 1 |
| | Number of generations | t | 10,000 |
| | Participation cost | g | 0.5 |
| | Investment of participation | c | 1 |
| | Participation benefit | r | 3 |
| | Strength of selection | w | 1 |
| | Mutation rate | u | 1e-10 |

**Table 4.2.** Model variables, parameters, and default of the parameter values.

| Imitation dynamics | Selection intensity | s | 0 ~ 1 |
|---|---|---|---|
| | Imitation probability | pr | 0 ~ 1 |
| | | | |
| Exploration dynamics | Exploration probabitlity | pe | 0 ~ 1 |
| | Mean for normally increment | mu | 0 ~ 1 |
| | Standard deviation for normally incrment | sigma | 0 ~ 1 |
| | | | |
| Created network | Number of nodes | n | 100 |
| | Connection probability | p | 0–1 |
| | Calculated social ties (from random graph) | t | 0–1 |

**Table 4.3.** Imitation, exploration, and created network characteristics of the parameter values.

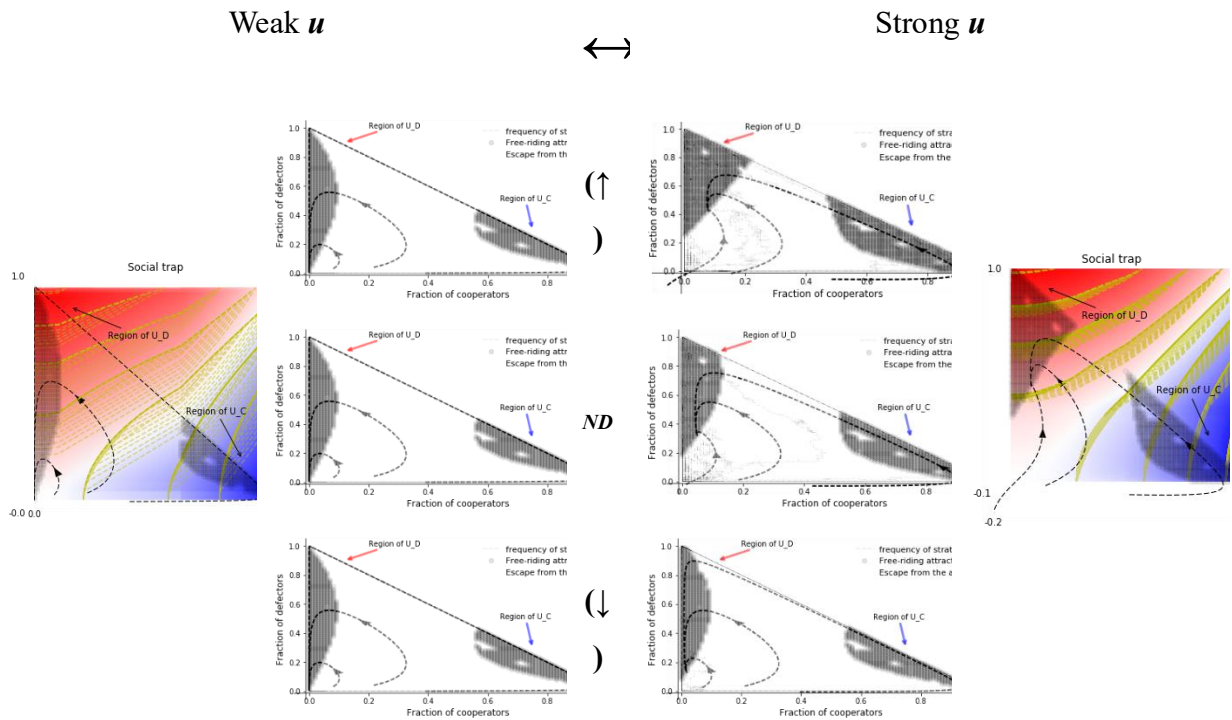

**Figure 5.** Cultural evolutionary dynamics using replicator-mutator, including network characteristics. In the left-hand plot, r = 3, c = 1, g = 0.5, and u = 1e-10. In the center-left plot (with network density = ND), r = 3, c = 1, g = 0.5, and u=1e-10 with upper ND = (0.9), bottom ND = (0.1). Center-right plot, r = 3, c = 1, g = 0.5, u=1e-10, upper ND = (0.9), and bottom ND = (0.1). Right-hand plot, r = 3, c = 1, g = 0.5, and u = 1e-3. The dotted lines show the homoclinic orbit for three assessment strategies (with different initial points). U_D denotes an unstable region of defection and cooperation.

In Figure 5, the exploration of the individual indicates its significance sensitivity, according to the exploratory trait of the mutation rate (Figure 5, left and right). However, the designated network density, as the individuals' social ties, can mediate its sensitivity. This means that when network density is high enough against the exploratory mutation rate, individuals are sensitized by mutation (Figure 5, center-upper panels) but the network density low, the phase portraits are not sufficient to be sensitive to changes in the mutation rate (Figure 5, center-bottom panels). This phenomenon of systemic sensitivity to external influence produces a more interesting evolutionary pattern.

# Model application

This proposed model for a public goods game represents a highly nonlinear system of replicator equations that can be analyzed with purely analytical means. For large incentive (r > 2), stable oscillations are observed, but where the cost for participation is too high (g > 1), no one will participate (Sigmund, 2010). From this, various combinations of time averages of the frequencies and payoffs of three strategies follow. It is found to be impossible to increase cooperation by increasing participation costs (or decreasing incentive), which favors defections and loners (Li, Jusup, Wang, Li, Shi, Podobnik., ... & Boccaletti, 2018). To promote cooperation, the incentive should be increased or he participation costs decreased, which would favor cooperation in the significant interest rate as well as in the experimental results (West, Griffin, & Gardner, 2007).

The simulation in the present model finds that the dynamics exhibit a wide variety of adaptive mechanisms that correspond to many different types of combinations, leading to various oscillations in the frequencies of the three strategies. The option to drop out of these dynamics depends on the mutation rate multiplied by network density as a social influence. In many societies, similar situations may occur, where small mutations are known to be a plausible risk in every network system and must have a marginal contribution to jeopardising the entire system (Helbing, 2013). Additional incentives attract larger participatory groups, but growth may inherently spell decline through mutations in any circumstance. However, the average effect of an individual's payoff remains the same, depending on the network characteristics, as if the possibility exists in this simulation, in relation to their social ties.

We investigated how the cooperation and defection changes with network characteristics with the involvement of the overall social heterogeneity (Santos, Santos, & Pacheco, 2008). For small ties among individuals, the heterogeneity remains low because the players only react slowly to social influence. On the other hand, as relationships grow, the dynamics develop rapidly enough to promote the social trap of defection (or free riding). Greater cooperation turns becomes additional competition at sites, which leads to a reduction in the overall network heterogeneity, given the results shown in this simulation. Reflection of the increase in heterogeneity of the pattern depends on the underlying societal organisation; much interaction is unable to eliminate the common trap and is not quickly eliminated by cooperators (Axelrod, 1986). Thus, the survival of cooperation relies on the individual's capacity to adjust to adverse ties, even when the rate of mutation (or systemic risk) is high. The results indicate that the simple adaptives of the social relations that are introduced here, coupled with the public goods game, account for a marginal contribution to the mitigating of systemic risk observed in realistic networks.

## Availability of data and material

All data and materials are our own. The materials and data used to support findings of this study are included within the Supplementary information file.

## Competing interests

The author declares that there is no competing interest regarding the publication of this article.

## Funding

This research was supported by the National Research Foundation of Korea (2016K2A9A1A02952017).

## Authors' contributions

CWP designed the model, tested it, and drafted the manuscript. The author read and approved the final manuscript.

## Acknowledgments

The author sincerely thanks to members of the Evolution and Ecology Program at International Institute for Applied Systems Analysis (IIASA) for their valuable support during the development of this article.